\documentclass[useAMS]{ws-mpla}
\usepackage{graphicx}
\usepackage[super]{cite}
\usepackage{multirow}
\usepackage{color}
\usepackage{enumerate}
\usepackage[table]{xcolor}
\usepackage{float}
\usepackage{comment}
\usepackage{enumerate}
\usepackage{adjustbox}
\begin{document}
\markboth{S.X. Yi}
{GW-Cherenkov radiation from photons}

\catchline{}{}{}{}{}

\title{On gravitational wave-Cherenkov radiation from photons when passing through diffused dark matters}
\author{\footnotesize Shu-Xu Yi}
\address{Department of Physics, The University of Hong Kong, Pokfulam Road, Hong Kong.\\
yishuxu@hku.hk}

\date{\today}
\maketitle
\begin{abstract}
Analogy to Cherenkov radiation, when a particle moves faster than the propagation velocity of gravitational wave in matter ($v>c_{\rm{g}}$), we expect gravitational wave-Cherenkov radiation (GWCR). In the situation that a photon travels across diffuse dark matters, the GWCR condition is always satisfied, photon will thence loss its energy all the path. This effect is long been ignored in the practice of astrophysics and cosmology, without justification with serious calculation. We study this effect for the first time, and shows that this energy loss time of the photon is far longer than the Hubble time, therefore justify the practice of ignoring this effect in astrophysics context.
\end{abstract}
\section{Introduction to gravitational wave-Cherenkov radiation}
Generally speaking, when a source of perturbation travels faster than the speed of propagation of that perturbation in medium, the front surfaces of influence from the source at different instances will coherently add up to form a shock wave. In the case that a stream of particles rams into another bulk of gas, if the velocity of the stream is larger than the sound speed in the gas, shock wave will occur; In the case of an electromagnetic field source, e.g. a charged particle, if the velocity of the particle is faster than the speed of light in the medium, an electromagnetic shock wave, i.e., Cherenkov radiation will arise.

When specific to gravitational perturbation, the source is a package of energy-momentum tensor and the propagation speed is the speed of gravitational wave (GW) $c_{\rm{g}}$. If some how the speed of the source can surpass $c_{\rm{g}}$, we would also expect the GW-shock wave, also known as the GW Cherenkov radiation (GWCR). However since the speed of GW is generally believed to be the speed of light in vacuum ($c_{\rm{g}}=c$), GWCR was thought impossible to occur in our ordinary physical world.  Some researchers \cite{1972NPhS..235....6L, 1972NPhS..236...79W, Schwartz2011} considered GWCR emitted by particles faster than $c$ (Tachyons). Since the existence of Tachyons is not wildly believed, those work receive limited attention. Another way to realize GWCR is to consider slower $c_{\rm{g}}$. Pioneered by ref.\,~\refcite{1980AnPhy.125...35C}, and recently ref.~\refcite{2001JHEP...09..023M} reconsidered the case of possible GWCR from ultra-energetic cosmic rays when $c_{\rm{g}}<c$ due to 4-D Lorentz violation (Rosen's bimetric theory of gravity in ref.\,~\refcite{1980AnPhy.125...35C}). They argued that in this case, GWCR sets a maximum travel time of a particle with given momentum. With the observation of energetic protons from $>10$ kpc, they set the upper bound of the difference between $c_g$ and $c$.

In fact, even in ordinary framework, $c_{\rm{g}}$ is less than $c$ when passing through matters due to dispersion. It is interesting to study the GWCR in diffused dark matters, which dominates mass of the universe in large scale.
For $c_{\rm{g}}$ in dark matters, we use the formula for dust~\cite{1980grg2.conf..393G}:
\begin{equation}
\frac{c^2}{c^2_{\rm{g}}}=1+\frac{4\pi G\rho}{3\omega_{\rm{g}}^2}, (\omega_{\rm{g}}\gg\sqrt{\frac{4\pi G\rho}{3}}).
\label{eqn:1}
\end{equation}
where $\omega_{\rm{g}}$ is the circular frequency of the GW.
For a particle with mass $m$, Lorentz factor $\Gamma$ and energy $E$, the range of the GWCR spectrum is limited by two conditions:
\begin{enumerate}
\item{the particle's velocity is larger than the phase velocity of GW at $\omega_{\rm{g}}$, which gives:}
\begin{equation}
\omega_{\rm{g}}<\sqrt{\Gamma\frac{4\pi G\rho}{3}}
\end{equation}
and
\item{the energy of each graviton $\hbar\omega_{\rm{g}}<E$, where $\hbar$ is the Plank constant divided by $2\pi$}.
\end{enumerate}
Since $\Gamma=E/mc^2$, for a given $E$, $m\rightarrow0$ gives $\Gamma\rightarrow\infty$. In this case, condition 1 is always satisfied and therefore the GWCR power is max.

Therefore, GWCR from photons traveling through diffused dark matters is the most important scenario to study this effect. This effect is completely ignored in the practice of astrophysics and cosmology, but without serious discussion. 
\section{Photons as sources of GWCR}
The energy spectrum of the Cherenkov Gravitational Wave Radiation is \cite{1994PhLB..336..362P}:
\begin{equation}
P(\omega_g)=\frac{G\omega_gE^2_{\rm{\gamma}}}{n^2c^5}(n^2-1)^2.\label{eqn:3}
\end{equation}
where $n\equiv c/c_{\rm{g}}$ is the refractive index of the GW.
Although this formula is derived with non-zero mass particles, we assume it also applies to photons~\cite{2012Optik.123..814G}. The energy losing rate of the photon is:
\begin{equation}
\frac{dE}{dt}=-KE_{\rm{\gamma}}^2,
\end{equation}
where
\begin{eqnarray}
K&\equiv&\frac{G}{c^5}\int_{\omega_{g,-}}^{\omega_{g,+}}\frac{\omega_g}{n^2}(n^2-1)^2d\omega_g\nonumber\\
&=&\frac{G}{2c^5}\int_{\omega_{g,-}}^{\omega_{g,+}}\frac{(n^2-1)^2}{n^2}d\omega_g^2.
\label{eqn:2}
\end{eqnarray}
From equation (\ref{eqn:1}) we know that:
\begin{equation}
\omega^2_{\rm{g}}=\frac{2\pi G\rho}{3(n-1)}.
\end{equation}
take above formula into the integration in equation (\ref{eqn:2}):
\begin{eqnarray}
K&=&\frac{G^2\rho\pi}{3c^5}\int^{n_+}_{n_-}\frac{(n^2-1)^2}{n^2(n-1)^2}dn\nonumber\\
&\approx&\frac{2G^2\rho\pi}{3c^5}(n_+-n_-)\nonumber\\
&<&\frac{2G^2\rho\pi}{3c^5}.
\end{eqnarray}
In the universe, galactic clusters are places with most ambient dark matters, where the average density of dark matters can be up to $\rho=10^{12}\,M_\odot/\rm{Mpc}^3$, where $M_\odot$ is the mass of the sun. Thus the upper limit of $K$ is:
\begin{equation}
K<4\times10^{-107}\,\rm{eV^{-1}s^{-1}}.
\end{equation}
The time life of the photon is:
\begin{eqnarray}
\tau&\equiv&E/\dot{E}\nonumber\\
&>&10^{106}\big(\frac{\text{eV}}{E}\big)\,\text{s}.
\end{eqnarray}
From the limit of photon-photon scattering process with the cosmology microwave back ground (CMB) and Extragalactic background light (EBL), the energy of photons cannot access $\sim10^{14}\text{eV}$. Therefore the life time of the photon under the GWCR is much longer than Hubble time, i.e., the age of our universe. As a conclusion, a lthough GWCR from photons when passing through diffuse dark matters is nonzero, it hardly leaves clues for its existence.

\section*{Acknowledgement}
The author appreciates the support from the department of physics, university of Hong Kong. The author thanks helpful discussion with Prof. Wu Kinwah and comments from anonymous reviewer.

\end{document}